\documentclass[fleqn,usenatbib]{mnras}

\usepackage{mathptmx}
\usepackage{txfonts}
\usepackage{float}
\usepackage{xcolor}
\usepackage[T1]{fontenc}

\DeclareRobustCommand{\VAN}[3]{#2}
\let\VANthebibliography\thebibliography
\def\thebibliography{\DeclareRobustCommand{\VAN}[3]{##3}\VANthebibliography}

\usepackage{graphicx}	
\usepackage{subcaption}

\title[Nustar observation of 4U 0114+65]{Nustar observation of the binary system 4U 0114+65}

\author[M. H. Abdallah et al.]{
Mohammed H. Abdallah,$^{1}$\thanks{E-mail: mohammed.abdallah@nriag.sci.eg}
Rasha M. Samir,$^{1}$
Denis A. Leahy,$^{2}$
Ashraf A. Shaker,$^{1}$
\\
$^{1}$Astronomy Department, National Research Institute of Astronomy and Geophysics (NRIAG), Cairo, Egypt\\
$^{2}$Department of Physics and Astronomy, University of Calgary, Calgary, Canada\\
}

\date{Accepted XXX. Received YYY; in original form ZZZ}

\pubyear{2023}

\begin{document}
\label{firstpage}
\pagerange{\pageref{firstpage}--\pageref{lastpage}}
\maketitle

\begin{abstract}

The high mass X-ray binary system 4U 0114+65 was observed by Nustar in October 2019, and by XMM-Newton in August 2015. Here we performed spectral and timing analysis of the Nustar observation, and carry out timing analysis on the XMM-Newton data. We measured the spin period of the neutron star from both observations and found a spin-up rate  $\dot p$ = $1.54$ $\pm$ $0.38$ $\times$ $10^{-6}$ $s$ $s^{-1}$.  During  the  Nustar observation two flares occured, one occured shortly after the start of the observation and the other near the end separated by a long period of low/quiescent- state. The large and sudden flares mostly resulted from accretion of Corotating Interaction Region (CIR) material. A common spectral model to HMXBs, powerlaw with high energy cutoff and absorption at low energy, gave a good fit to both flaring and quiescent states.  A flourescent  iron line was not required in fitting any of the states. On the other hand, very tentative evidence of  Cyclotron Resonant Scattering Feature  (CRSF) at $\sim$ 17 keV was found during fitting using $cyclabs$ model, however fitting improvement was not significant enough to confirm its detection, plus a very narrow width (< 1 keV) was obtained for the line and its first harmonic. Visual inspection of the spectra showed  a  deficiency of emission near the expected first and second harmonic. Another important feature visually noticed in the spectra is the presence of hard tail above 50 keV. This could be explained by the shocked material bounding  the  CIR.

\end{abstract}




\section{Introduction}

X-ray emission in supergiant X-ray binaries containing a neutron star is produced via accretion of matter by the compact object from its optical companion. This material transfer can take place through Roche-lobe overflow or the capture of stellar wind. The source 4U 0114+65 is generally believed to be a wind-fed supergiant X-ray binary. The system was discovered in the late seventies during the SAS-3  Galactic  survey (Dower and Kelley 1977), and shortly after that its companion LSI +65$^{\circ}$010 was identified (Margon and Bradt, 1977). Since its discovery many of the physical parameters of the system remain debatable.  The binary nature of the system was first confirmed by the finding of orbital modulations of optical radial velocity measurements (Crampton et al. 1985). Using RXTE/ASM observations, variability of X-ray emission over slightly different orbital period was obtained (Corbet et al. 1999). With new optical spectroscopic observations and the reanalysis of the X-ray observations a consistent binary period was determined. The reported values of the orbital period of the system are 11.5983 +/- 0.0006 day with non-zero eccentricity of 0.18 +/- 0.05 from optical observations (Grundstrom et al. 2007) and 11.599 +/- 0.005 day from X-ray observations (Wen et al. 2006). The binary system shows characteristics of both Be and supergiant X-ray binary (Koenigsberger et al. 1983, Crampton et al. 1985). Reig et al. 1996, reclassified the optical star as a luminous B1 Ia supergiant at a distance of $\sim$ 7.2 kpc. This classification was further confirmed using near infrared observations (Ashok et al 2006). 

\vspace{0.3 cm}

X-ray emission from 4U 0114+65 show variabilities over a wide range of time scales. The system shows flares that last for few ksec , occuring randomly with peak intensity of $\sim 15-20$ times that of the persistent low state emission (Masetti et al. 2006; and Mukherjee \& Paul 2006). Masetti et al. 2006 suggested that these flares are caused by random inhomogeneities in the accretion flow onto the neutron star and thus accretion flux. X-ray emission from the neutron star also  shows  short-term variabilities (X-ray flickering) on a time scale of minutes (Koenigsberger et al. 1983). These flickerings were previously reported as the rotation period of the neutron star by Koenigsberger et al. (1983); and Yamauchi et al. (1990) with periods of 894 and 850 seconds respectively. These short-term variations were also reported with different periods at 2000 s (Apparao et al. 1991) and 710 s to 1470 s (Farrell et al. 2008) and mostly  explained  by clumping of the accreted stellar winds or due to feedback between the radiation pressure and the accreted material (Farrell et al. 2008). Using observations extending over seven years taken by EXOSAT, GINGA, and ROSAT Finley et al. (1992) found a persistent periodic modulations of period 2.78 h. This periodicity was confirmed as the true pulse period of the neutron star (Corbet et al. 1999; Hall et al. 2000) using RXTE/ASM and RXTE/PCA observations respectively.  Continued  X-ray observations of the system 4U 0114+65 showed that the pulsation  persists  up to high energy  X-rays  of 80 keV (Bonning \& Falanga 2005) and the rotation of the neutron star is getting faster over time (Bonning \& Falanga 2005; Farrell et al. 2008; Wang 2011; and Hu et al. 2017). The neutron star in 4U 0114+65 is one of the slowest rotating X-ray pulsars. Theoretical models to explain how the neutron star reached such very slow rotation include the assumption by  Li \& van den Heuvel (1999). They suggested that the neutron star was born as magnetar with initial magnetic field in excess of $10^{14} G$. Such high magnetic field will naturally increase the equilibrium period reached during the propeller phase. Ikhsanov (2007), on the other hand didn't require such high magnetic field by introducing an extra phase (subsonic propeller phase) on the evolutionary track of the neutron stars in massive binary systems. In addition to the previous X-ray emission variations, 4U 0114+65 exhibits stable super-orbital modulations with a period of $\sim 30.7$ day (Farrell et al. 2006; Wen et al. 2006; and Kotze \& Charles 2012). The cause of this modulation is still unclear.  

\vspace{0.3 cm}

The continuum X-ray spectrum of 4U 0114+65 is typically fitted with the classical phenomenological model of accreting pulsars, namely absorbed power law with a high energy exponential cut-off (White et al. 1983).  The typical parameters for the system are photon-index $\Gamma \sim 1$ , cutoff energy $E_{cut} \sim 8$ keV , folding energy $E_{fold} \sim 20$ keV with hydrogen column density  showing wide range of values $2-15$  $\times$  $10^{22} cm^{-2}$  (Hall et al. 2000; Masetti et al. 2006; Farrell et al. 2008). Inclusion of  an  iron K$\alpha$ line was sometimes found to be needed and other times not required. The emission line is preferentially detected during the persistent quiescent state of the system (Masetti et al. 2006) or during the low state of the pulse profile (Hall et al. 2000). On the other hand, Mukherjee (2006) found appreciable presence of the iron line during both extended low state and the flaring state with comparable equivalent width in both cases. Yet in other times inclusion of Fe K$\alpha$ line didn't improve the fits (Farrell et al. 2008). Bonning \& Falanga (2005) reported detection of a Cyclotron Resonant Scattering Feature (CRSF) at $\sim$ 22 keV, however such absorption feature was not confirmed in any later work.

\vspace{0.3 cm}

In this paper we carry out timing and spectral analysis of  a  Nustar observation of 4U 0114+65. In the timing analysis we aim to examine pulse periodicity.  For that purpose we coded the computations for periodicity searching of Lomb-Scargle periodogram (LSP) (Lomb 1976; Scargle 1982), Phase Dispersion Minimization (PDM) method (Stellingwerf 1978), and Epoch-Folding (EF) technique (Leahy et al. 1983; Leahy 1987).  To further check our computations, and to measure the drivative of the pulse period, we reanalyzed  the  XMM-Newton observation of the system and compare our results with those given in Sanjurjo-Ferrr{\'\i}n et al. (2017). Observations and data reduction are  briefly  given in section (2). Timing and spectral analysis, including our results are detailed in section (3). We then discuss our results and  conclude  the paper in section (4).

\begin{figure}
	\includegraphics[height=8cm, width=10cm]{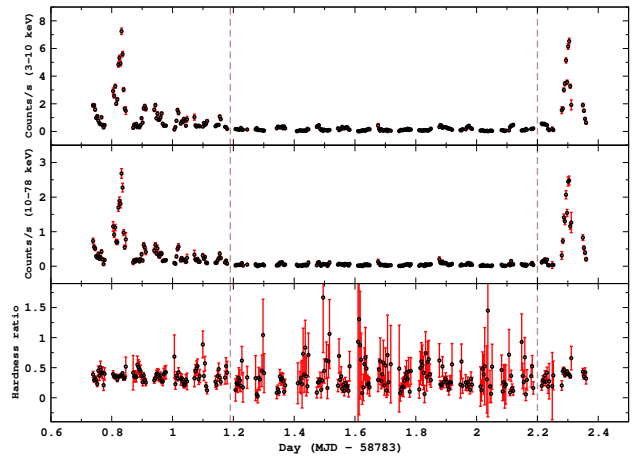}
    \caption{Nustar  background  subtracted light curves of 4U 0114+65 with time binning of 300-sec: (top panel) 3-10 keV band; (middle panel) 10-78 keV band; (bottom panel) Hardness ratio. Vertical dashed lines separate the plots into three sections: Flare A (left), Quiescent state (middle), and Flare B (right).}
    \label{fig:lightcurve_nustar}
\end{figure}

\begin{figure}
	\includegraphics[height=8cm, width=10cm]{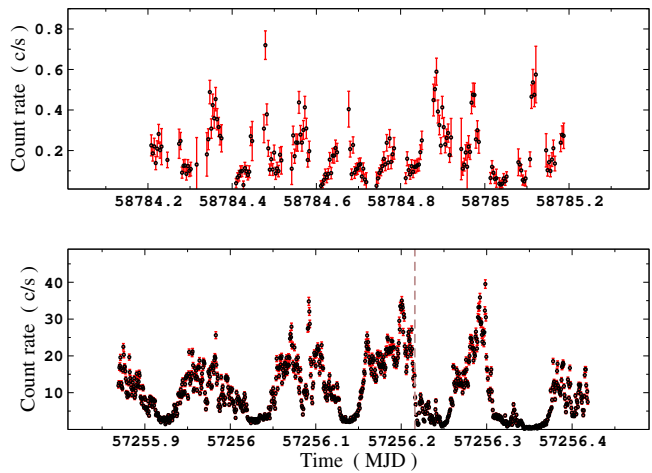}
    \caption{Top panel: Nustar  background  subtracted light curve of the quiescent state in the 3-78 keV band with time binning of 300-sec. Bottom panel: XMM-Newton  background  subtracted light curve of 4U 0114+65 in the 3-10 keV with time binning of 50-sec, vertical-dashed line separates the first three (T1-section) and the last two (T2-section) pulses. }
    \label{fig:lightcurve_nuxmm}
\end{figure}

\section{Observations and data reduction}

\subsection{Nustar Observation}

The Nuclear Spectroscopic Telescope Array (Nustar) (Harrison et al. 2013), consists of two focal plane modules (FPMA \& FPMB) operating in the hard X-ray band (3-78 keV).  Observation  of 4U 0114+65 was performed on 2019 October 27 (MJD  	58783.73; observation ID 30501016002) for exposure time of 69.5 ksec. Data were processed using Nustar Data Analysis Software (Nustardas version 2.0.0) distributed with HEASOFT 6.28 release and calibration files (CALDB v20210104). We used $\emph{nupipeline}$ task to create calibrated event files from both detector modules. Filtering and screening was performed using $\emph{nufilter}$ \& $\emph{nuscreen}$ tasks. We then used $\emph{nuproduct}$ script for the production of scientific products (image, lightcurve, spectra, and response files). For lightcurve/spectra of the source we extracted the data from a circular region centered on the source with a radius of 50 arc-sec, corresponding to encircled energy fraction of $\sim 70\%$ (An et al. 2014).  Similarly for background products a circular region of the same size in a blank area on each of the detectors was used. Barycentric corrections were accounted for using source position RA = 19.51125 and DEC = +65.29161. 

\subsection{XMM-Newton Observation}

XMM-Newton mission (Jansen et al. 2001), carries three co-aligned X-ray telescopes. Two of the telescopes are equipped with gratings diverting half of the incident X-ray flux for spectroscopic observations, leaving the other half for the imaging detector "EPIC-mos camera". In the third telescope all of the incident flux reaches the imaging camera EPIC-pn. In the current work, we use only data obtained by the EPIC-pn camera. 4U 0114+65 was observed by XMM-Newton on 2015 August 21 (MJD 57255.85 observation ID 0764650101) for a total duration of 49 ksec. The EPIC-pn camera was operating in the small window mode. Observation data files (ODFs) were reduced by XMM-Newton Science Analysis Software (SAS version 18.0), with the latest Current Calibration Files (CCFs). We processed the pn data by running $\emph{epchain}$ routine. The event file was checked and found to be free of high background flaring periods. We used the SAS tool $\emph{barycen}$ to correct the arrival time of events to the solar system barycentre. The source emission was taken from a circular region centered on the source and has a radius of 30 arc-sec  , corresponding to encircled energy fraction of $\sim 90\%$ (Aschenbach et al. 2000). The  background was taken from a circular region of the same size in a blank area. We used the SAS task $\emph{evselect}$ to extract light curves for the source and background and the source light curve was then corrected and cleaned using the SAS task $\emph{epiclccorr}$.

\section{Analysis and Results}
\subsection{Light curves}
We extracted the light curves from both focal plane modules (FPM A/B) of Nustar observations of 4U 0114+65 in the energy bands 3-10 keV \& 10-78 keV with bin-size of 300 s. The subtracted light curves were then add using Xronos task $\emph{lcmath}$. The light curves and the hardness ratio are shown in figure (1). As clear from the figure there are flares in the begining and end of  the observation  and a quiescent state in between. The main flare in both lasted for about 4-6 ksec (not including pre/post-activity in each). In subsequent analysis we divided the data into three sections separated in figure (1) by vertical-dashed lines. The first section is the flare at the begining of the observation (from now on referred to as Flare-A), the second section is the quiescent state, and the last section is the flare at the end of the observation (Flare-B). In the energy band 3-78 keV, the count rate of the flare's peak are $\sim 48$ and $13$ times of the mean and maximum count rates during the quiescent state respectively. A close view of the light curve during the quiescent state in the 3-78 keV band is shown in the top panel of figure (2).      

\vspace{0.3 cm}

XMM-Newton observation of 4U 0114+65 covers about five pulses of the neutron star. This observation was analyzed before (Sanjurjo-Ferrr{\'\i}n et al. 2017). They used only pn-data for timing analysis. We follow their procedure in timing analysis, in which they divided X-ray light curve into two sections: T1-section coveing the first three pulses , and T2-section for the last two pulses. For timing analysis they used only T1-section of the observation above 8 keV to avoid absorption effect. The 3-10 keV light curve of XMM-Newton pn-data with bin-size of 50 sec is shown in lower panel of figure (2), with vertical dashed line separating T1 \& T2 sections of the observation.

\subsection{Timing Analysis}
For timing analysis we coded the computations  of  Lomb-Scargle periodogram (LSP) (Lomb 1976; Scargle 1982), Phase Dispersion Minimization (PDM) method (Stellingwerf 1978)  , and Epoch-Folding (EF) technique (Leahy et al. 1983; Leahy 1987).  In the Nustar observation, flares and the large flux variability during pre/post flare will affect the search for pulse periodicity, thus we used only the quiescent data in our computations of the pulse period. The quiescent state extend over about nine pulses of the neutron star with an average coverage of each by about 56\%. We extracted light curves from FPMA and FPMB detectors in the energy rage 3-78 keV with time bin-size of 10-sec. The large noise in the individual light curves and possibly the incomplete coverage of the individual pulses resulted in a very noisy result of the $\theta$-statistics curve of PDM method. Thus both light curves were added using $lcmath$ task for the computation of periodicity. In PDM computations, we used number of bins $N_{b}$ of 200 and covering number $N_{c}$ of 2 and searched for periodicity in the range from 5 to 15 ksec with time resolution of 20-sec. The Lomb-Scargle periodogram gave a consistent results of both the individual light curves and the added light curve. In LSP, the searched frequencies are between 2 cycles/day to 400 cycles/day with resolution that varies appropriately within the searched frequencies. Results of the computations are plotted  in figure (3).

\vspace{0.3 cm}

Similarly, for the XMM-Newton observation of 4U 0114+65, we extracted the light curve from pn-camera with time binning of 10-sec for timing analysis. We used data in the band 3-10 keV to avoid absorption effects in the soft band. We found this choice is sufficient and there  was  no need to restrict the analysis in the 8-10 keV band. In PDM computations we used $N_{b}$ of 100 and $N_{c}$ of 2 computing $\theta$-statistics in the range from 5 to 15 ksec with time step of 5-sec. Computations of Lomb-Scargle power were also done similar to that of the Nustar data. We run the computations using both methods for T1-section of the observation as well as the whole of the observation (i.e. T1 + T2). Figure (4) show the results of the computations. The periods obtained using either PDM or LSP are given in table (1).These periods and the associated uncertainities were obtained in each case by bootstrap sampling of a 1000 random samples extracted from and with similar size of the original light-curves. The sampling was done with replacement (i.e. repetition is allowed). For each sample, we performaed the LSP and PDM computations over shorter period range around the best value obtained from the original light curves to reduce computation time. The distributions of the best periods obtained are shown in figure (5), and table (2) gives the details of the computations performed ( period range, resolution, and relevant comment in each case). The quoted values for PDM and LSP in table (1) are the mean and the standard deviation of the bootstrap distribution. 

\vspace{0.3 cm}

 To further check  the periods, we performed the computation of Epoch-Folding (EF) technique (Leahy et al. 1983; Leahy 1987) for all data. For XMM-Newton observation we folded the light curve into 100 bins and the S-statistics (equation 12 in Leahy et al. 1983) were computed for each trail period from 5000 to 15000 seconds with time step of 1-second. To determine the period a chi-square fitting with the appropriate $sinc^{2}$ function (Leahy 1987) was performed. The fitting was performed within 300-sec around the test period (i.e. $|P^{\prime}-P|$ $\leq$ 150  see Leahy 1987 for definition of $|P^{\prime}-P|$ ). The standard deviation was obtained using equation (6a) in Leahy (1987). The fitting of T1-section of XMM-Newton observation did not result in a well constrained fit (varies with the change of $|P^{\prime}-P|$) and mostly resulted in higher values for the spin period around (9360 to 9420 sec). Similar steps were taken in determining the periodicity in Nustar observation, however the light curve was folded into 20 bins to reduce the noise and better constrain the period. The results of periodicity using EF technique are shown in figure (6) and given in table (1).  

For the XMM-Newton observation, as clear from the results that LSP gave different values of the period for T1-data and the whole of the observation. This difference is caused mainly due to pulse profile change between T1 \& T2 sections of the observation as noted by Sanjurjo-Ferrr{\'\i}n et al. (2017). On the other hand Phase folding techniques (i.e. PDM \& EF) seem to be insensitive to such profile change in the pulses and gave a value for the period that is consistent with the weighted average value obtained by Sanjurjo-Ferrr{\'\i}n et al. (2017) using various methods. LSP assumes a sinusoidal model for the data, while phase folding methods (PDM \& EF) have no prior assumption on the shape of the pulses. The convolution of the assumed sinusoidal model with the data in LSP resulted in a Gaussian distribution of the periods obtained using bootstrap sampling. In PDM computations, the absense of any prior assumption of the pulse profile and the noisy nature of the computions resulted in a not Gaussian-like distribution of bootstrap sampling results. The extreme cases are the distribution obtained in case of XMM-Newon observation, where only few bins shows the highest probability. This mostly occured due to the skewness apparent in the PDM results around the best period (Fig. 4, right panels). The T1-section PDM results show higher degree of skewness, resulting in biasing most of bootstrap sampling results towards a lower value around 9280 sec. This skewness is also apparent in EF-computation of T1-section of the observation and it makes it difficult to obtain a converg fit. 

To summarize this section, the timing analysis of Nustar obsrvation is of high success where results from all methods agree with each other.  The average value obtained using all three methods is ( 9143 $\pm$ 34 sec. ). For the XMM-Newton observation, LSP failed in finding the spin period of the neutron star as we obtained completely different values for periods  obtained from T1-section and all of the observation. Phase folding methods (PDM \& EF) did not show such large difference in the values of the spin period obtained from both data (i.e. T1-section and all of the observation) : PDM results for both data agree with each other within uncertainty and both agree with EF results of all XMM-Newton observation. Although we could not obtain a converge fit to EF results for T1-section, all trial fits resulted in a slightly higher values mainly caused by the high degree of skewness in phase folding results for T1-section. It is clear that phase folding results for T1-section of the data show a large degree of skewness, thus making these results much less reliable. Given the much difficulity in constraining the neutron star spin period during XMM-Newton observation, we find that PDM results obtained for all XMM-Newton observation ( 9346 $\pm$ 37 sec. ) is the most reliable and most conservative value for the neutron star spin. The average spin up rate between observations, using the above mentioned periods, is found to be $\dot p$ = $1.54$ $\pm$ $0.38$ $\times$ $10^{-6}$ $s$ $s^{-1}$.

\begin{table}
	\centering
	\caption{ Pulse period measurements in seconds using Lomb-Scargle periodogram (LSP), Phase Dispersion Minimization (PDM), and Epoch Folding (EF).}
	\label{tab:example_table}
	\begin{tabular}{llll} 
		\hline
		Observation & LSP & PDM & EF\\
		\hline
		XMM-Newton  & 9178 $\pm$ 19  & 9346 $\pm$ 37 & 9315 $\pm$ 7 \\
		(All observation) & & & \\
		 & & & \\
		XMM-Newton  & 9545 $\pm$ 25  & 9312 $\pm$ 46 &  \\
		(T1-section) & & & \\
		 & & & \\
		Nustar  & 9136 $\pm$ 32 & 9149 $\pm$ 51 & 9144 $\pm$ 20 \\
		(Quiescent data) & & & \\
		 & & & \\
		\hline
	\end{tabular}
\end{table}

\begin{table}
	\centering
	\caption{ Details of bootstrap sampling computaions.}
	\label{tab:example_table}
	\begin{tabular}{llll} 
		\hline
		Observation and method & range & resolution & Other comments\\
		\hline
		XMM-Newton  &  &   &   \\
		(All observation) &  &   &   \\
		PDM & 9.2-9.5 ksec & 1 sec & $N_{b}$ = 50   \\
		 & & & $N_{c}$ = 2\\
		LSP ( $day^{-1}$ ) & 8.5 - 10.2  & 0.0002  &  \\
		\hline
		XMM-Newton &    &   &   \\
		(T1) &    &   &   \\
		PDM & 9.2-9.5 ksec & 1 sec & $N_{b}$ = 50  \\
		 & & & $N_{c}$ = 2\\
		LSP ( $day^{-1}$ ) & 8.5 - 10.2  & 0.0002  &  \\
		\hline
		Nustar &    &   &   \\
		PDM & 8.5-9.7 ksec & 1 sec & $N_{b}$ = 200   \\
		 & & & $N_{c}$ = 2\\
		LSP ( $day^{-1}$ ) & 9 - 10.2 & 0.0002  &  \\
		\hline
	\end{tabular}
\end{table}

\begin{table}
	\centering
	\caption{The spectral parameters with 1-$\sigma$ error of Nustar spectra of different emission states and the total spectra.}
	\label{tab:example_table}
	\begin{tabular}{lcccc} 
		\hline
		parameter & Flare A & Quiescent & Flare B & Total \\
		\hline
		$N_{H}$ ($10^{22}$ $cm^{-2}$) & $3.90^{+0.92}_{-0.64}$  & $5.09^{+1.78}_{-1.85}$ & $4.61^{+1.11}_{-1.15}$ & $4.48^{+0.51}_{-0.64}$ \\
		 &   &  &  & \\
		$\Gamma$ & $1.65^{+0.09}_{-0.07}$  & $2.25^{+0.20}_{-0.25}$ & $1.59^{+0.10}_{-0.11}$ & $1.75^{+0.05}_{-0.07}$ \\ 
		 &   &  &  & \\
		$E_{cut}$ & $7.60^{+0.44}_{-0.58}$  & $7.47^{+2.25}_{-1.35}$ & $7.87^{+0.96}_{-0.75}$ & $7.76^{+0.38}_{-0.39}$ \\ 
		 &   &  &  & \\
		$E_{fold}$ & $25.8^{+5.2}_{-2.8}$  & $28.4^{+25.8}_{-10.8}$ & $25.5^{+5.5}_{-4.1}$ & $28.5^{+3.5}_{-3.5}$ \\
		 &   &  &  & \\
		 FPMA PL norm  ($10^{-3}$)& $5.54^{+1.16}_{-0.70}$  & $2.01^{+0.99}_{-0.76}$ & $8.67^{+2.06}_{-1.80}$ & $3.46^{+0.36}_{-0.43}$ \\
		 &   &  &  & \\
		 FPMB PL norm  ($10^{-3}$)& $5.79^{+1.21}_{-0.73}$  & $2.14^{+1.04}_{-0.65}$ & $8.82^{+2.09}_{-1.83}$ & $3.57^{+0.37}_{-0.44}$ \\
		 &   &  &  & \\
		 $\chi^{2}$ (dof) & 540 (531)  & 144 (133) & 326 (313) &  640 (661) \\
		 &   &  &  & \\
		\hline
	\end{tabular}
\end{table}

\begin{figure}
	\includegraphics[height=8cm, width=8cm]{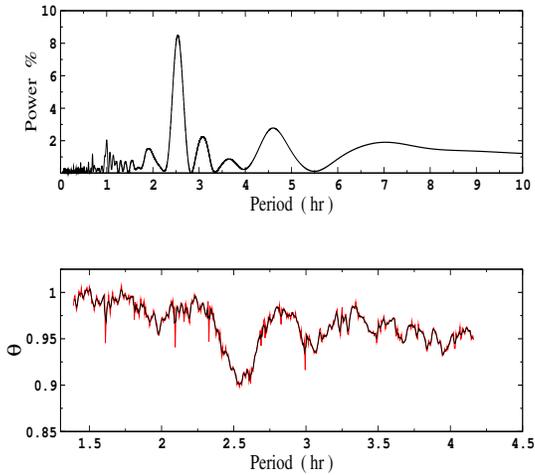}
    \caption{Top-panel: Lomb-Scargle power, bottom-panel: the $\theta$ statistic of PDM method (red), overploted with a smoothed 3-point running average (black). Both plots are for the quiescent state of the Nustar observation of 4U 0114+65.}
    \label{fig:period_nustar}
\end{figure}

\begin{figure}
	\includegraphics[height=6cm, width=8cm]{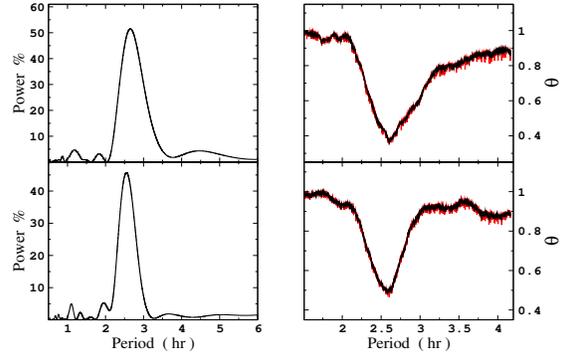}
    \caption{Left-panels: Lomb-Scargle power, right-panels: the $\theta$ statistic of PDM method (red), overploted with a smoothed 3-point running average (black). Top plots are for T1 section of XMM-Newton observation of 4U0114+65, and bottom plots are for the whole XMM-Newton observation.}
    \label{fig:period_xmm}
\end{figure}

\begin{figure}
     \centering
     \begin{subfigure}[b]{4cm}
         \centering
         \includegraphics[height=4cm, width=4cm]{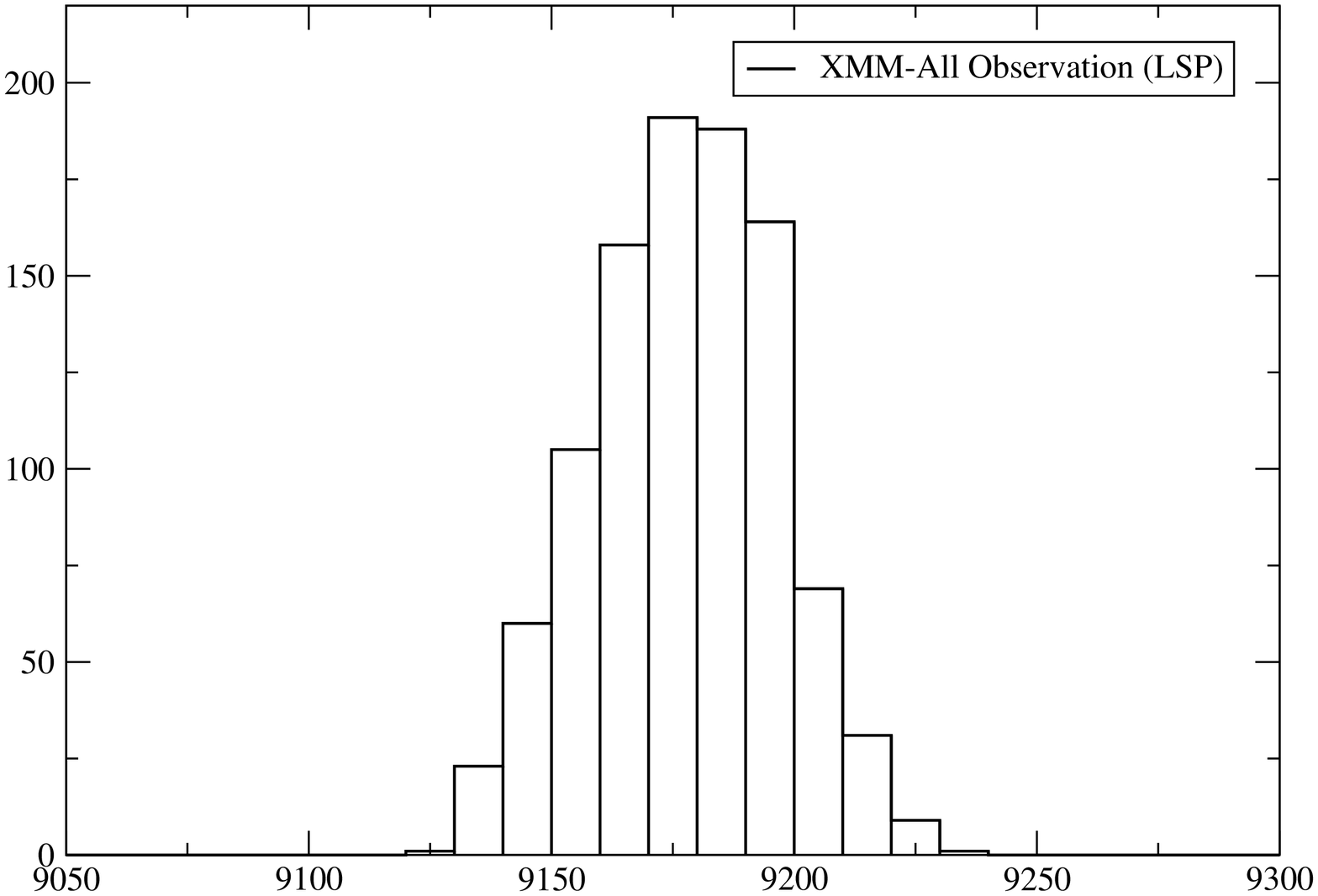}
     \end{subfigure}
     \hfill
     \begin{subfigure}[b]{4cm}
         \centering
         \includegraphics[height=4cm, width=4cm]{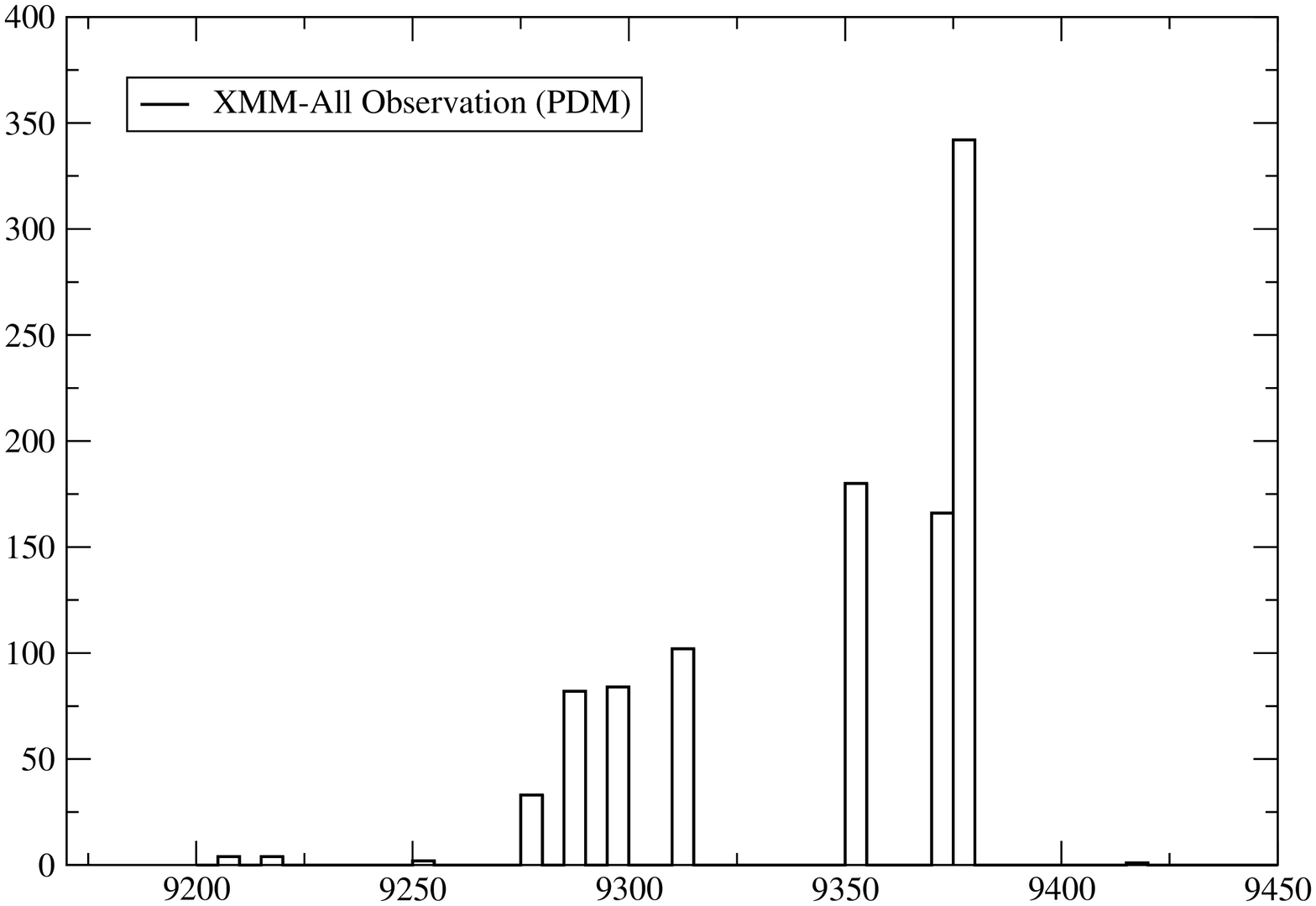}
     \end{subfigure}
     \vfill
     \begin{subfigure}[b]{4cm}
         \centering
         \includegraphics[height=4cm, width=4cm]{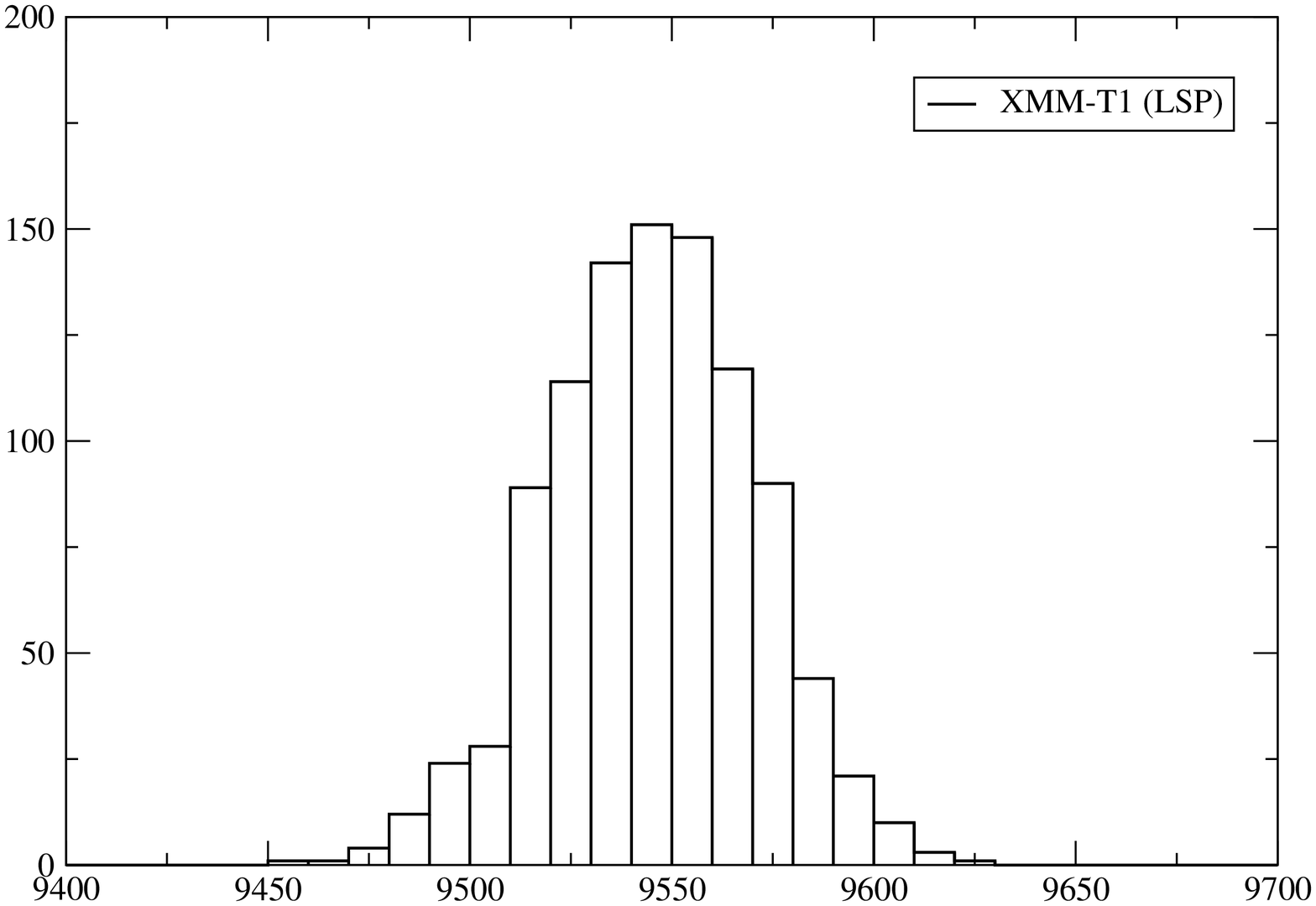}
     \end{subfigure}
     \hfill
     \begin{subfigure}[b]{4cm}
         \centering
         \includegraphics[height=4cm, width=4cm]{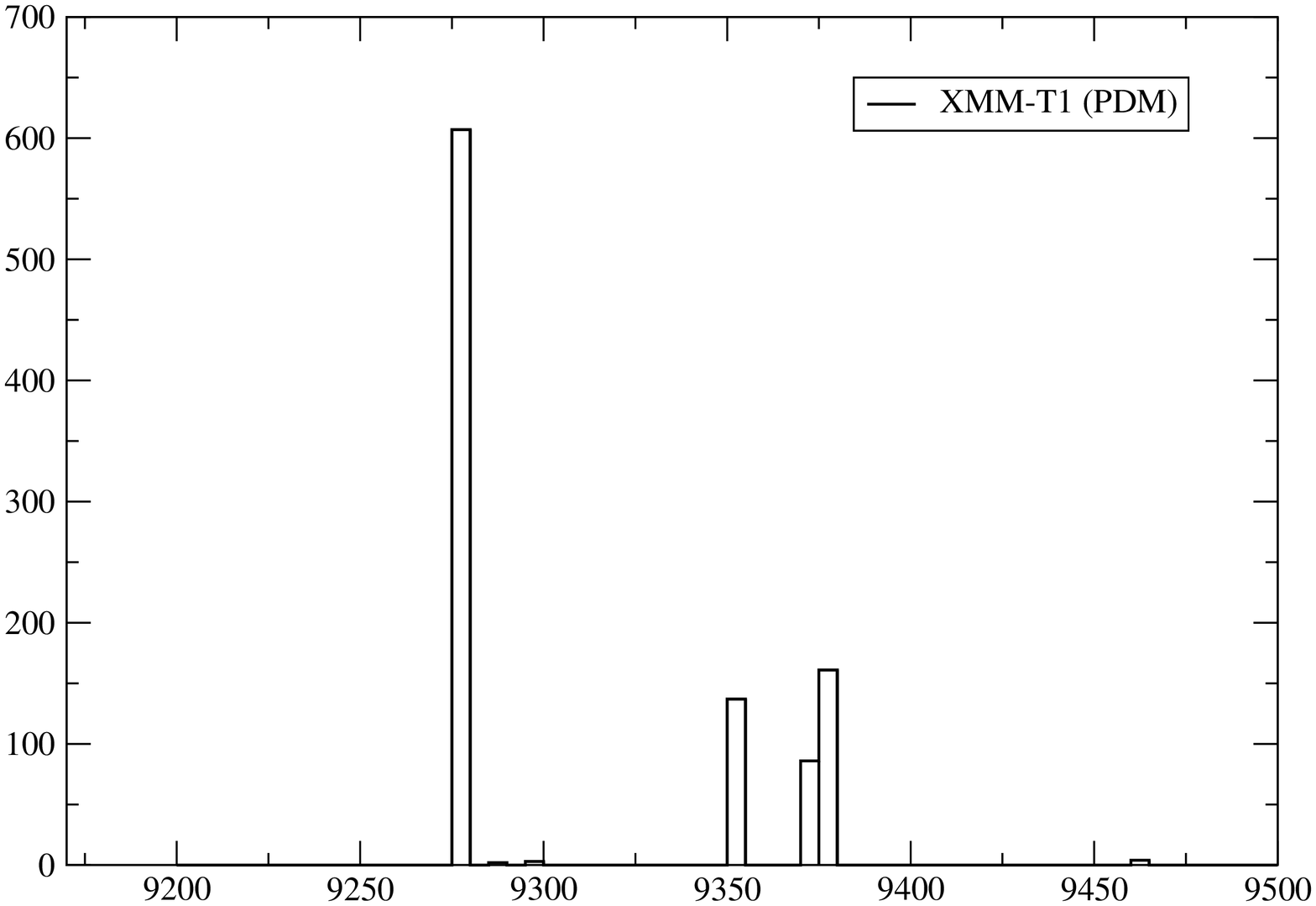}
     \end{subfigure}
     \vfill
     \begin{subfigure}[b]{4cm}
         \centering
         \includegraphics[height=4cm, width=4cm]{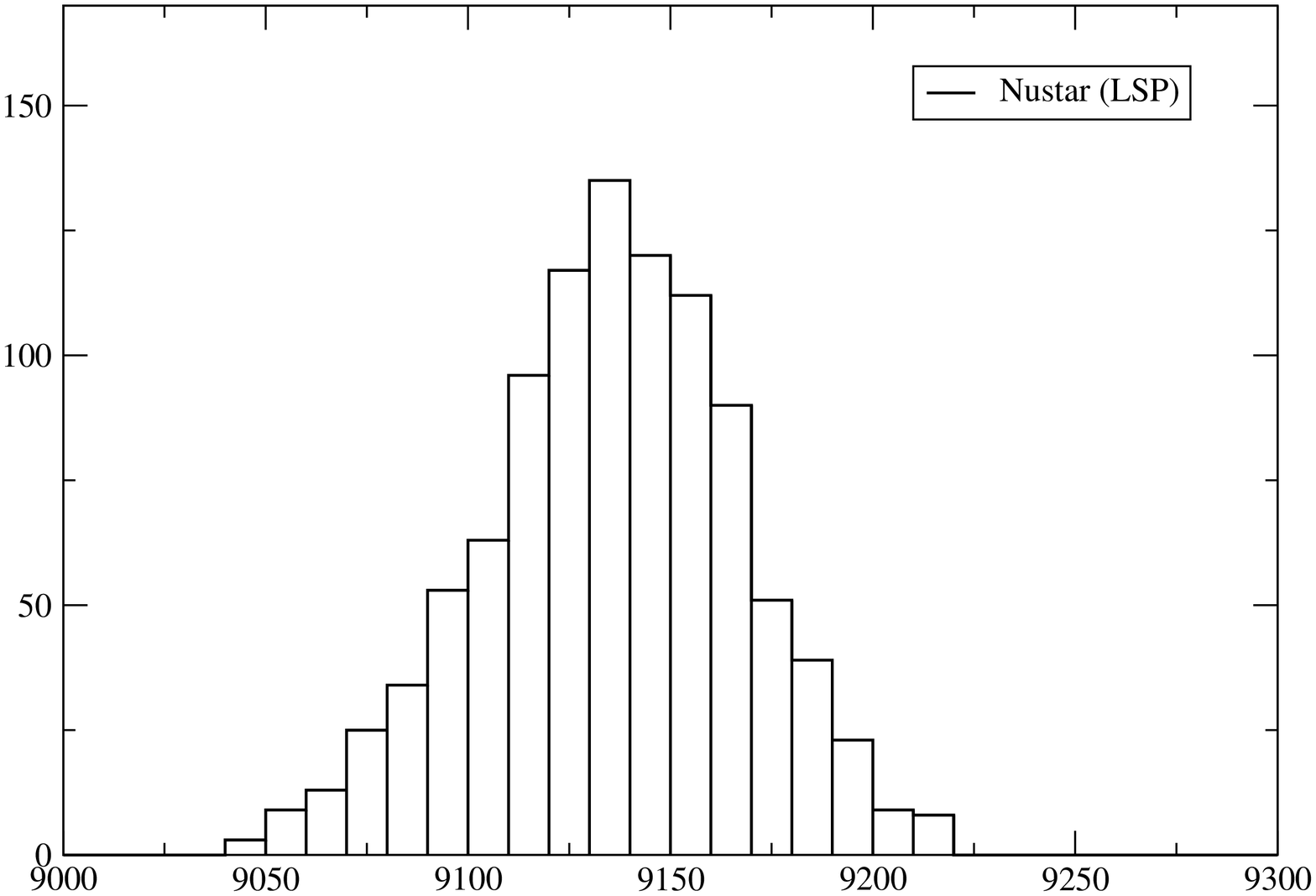}
     \end{subfigure}
     \hfill
     \begin{subfigure}[b]{4cm}
         \centering
         \includegraphics[height=4cm, width=4cm]{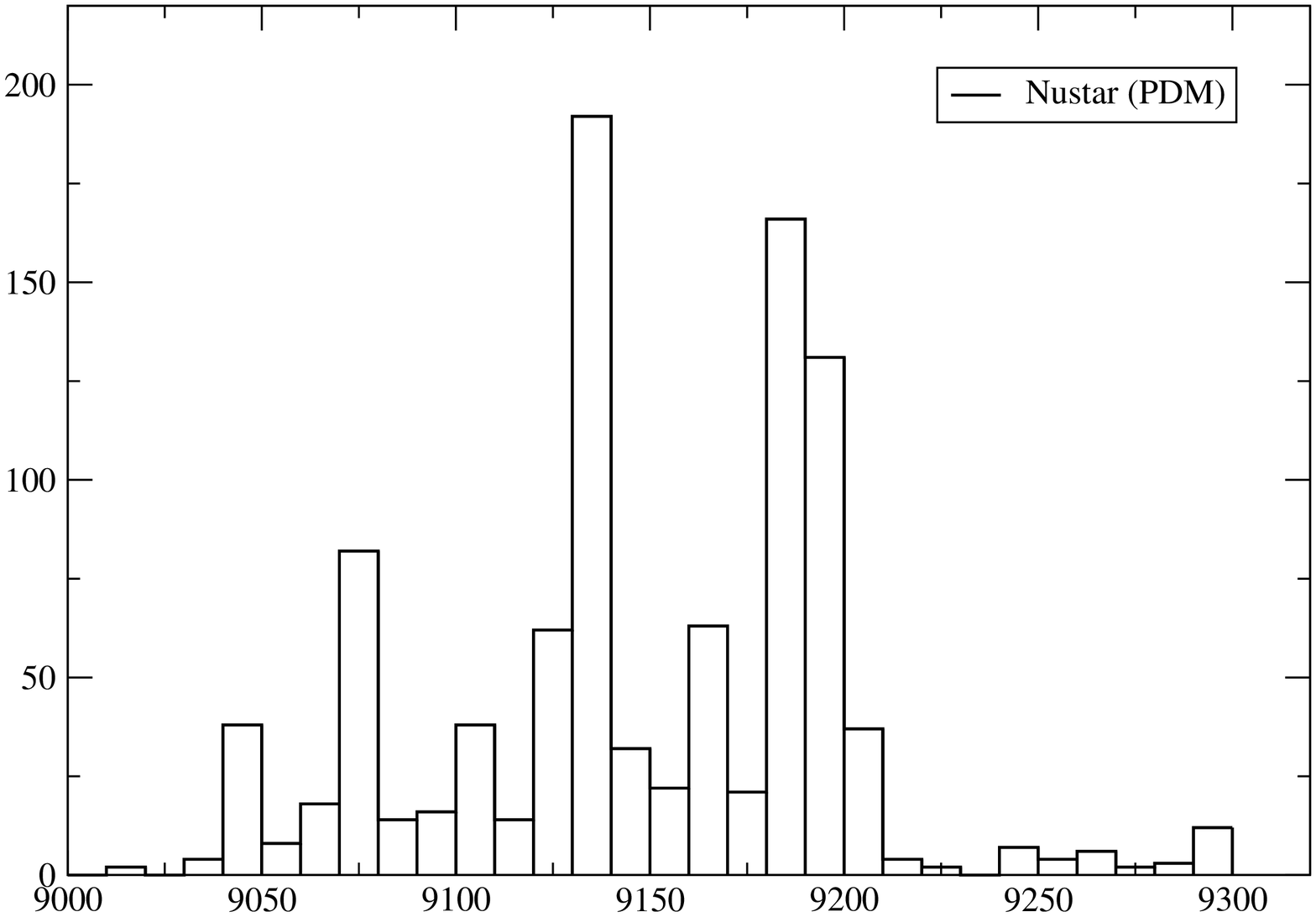}
     \end{subfigure}
     \caption{ Period distribution for 1000 random samples in each case. Left: LSP results , Right: PDM results , Top: XMM-Newton all observation , Middle: XMM-Newton T1 data , Bottom: Nustar observation. X-axis: period in seconds , Y-axis: number of samples}
        \label{fig:three graphs}
\end{figure}

\begin{figure}
	\includegraphics[height=7cm, width=9cm]{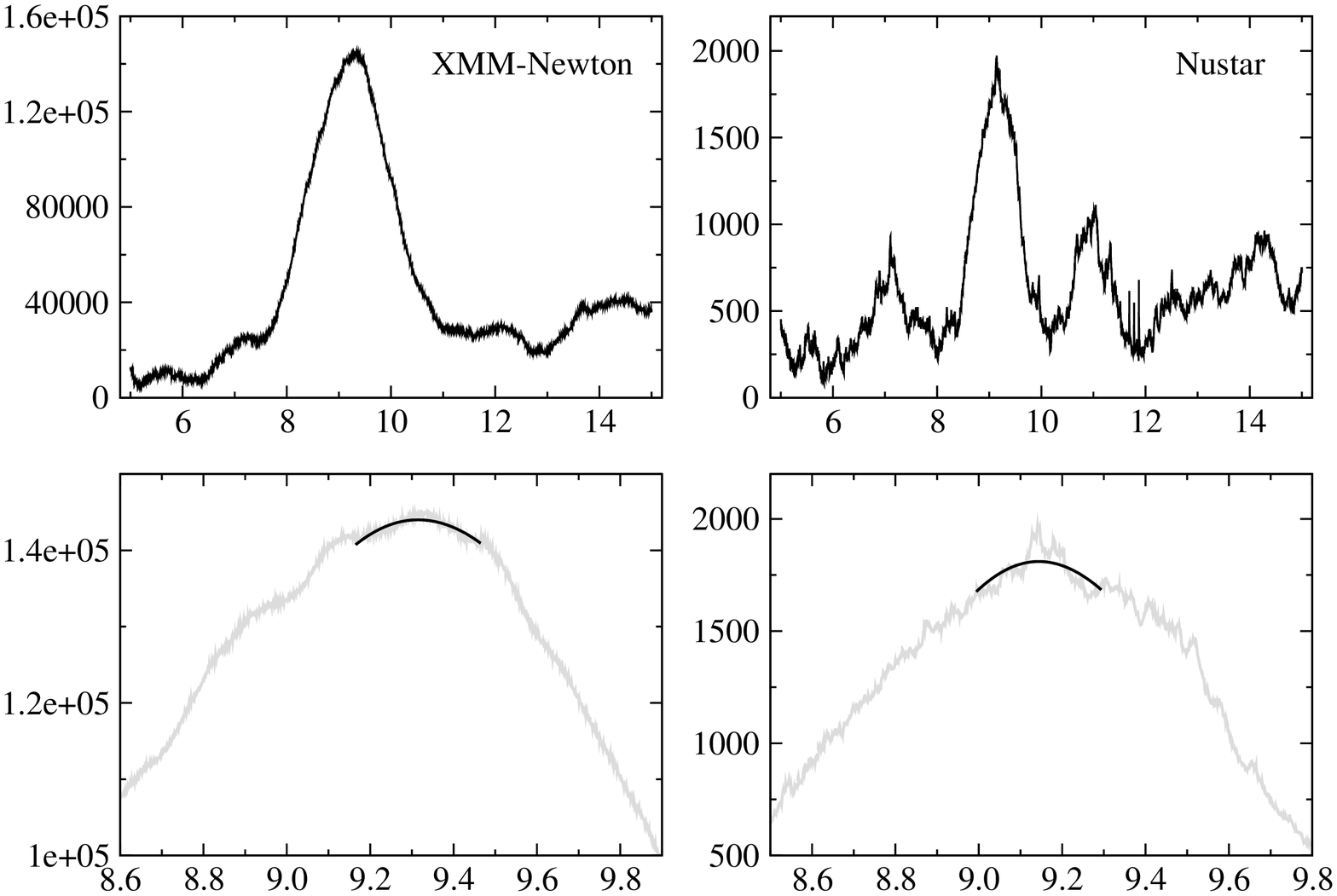}
    \caption{ Epoch folding results, Left-panels: for XMM-Newton all observation, Right-panels: for Nustar observation. Top-panels: the S-statistics as computed using Equation (12) in Leahy et al. (1983), Bottom-panels: zoomed view showing the S-statistics and the best fit $sinc^{2}$ function as given by equation (4) in Leahy (1987). X-axis: The period in ksec, Y-axis: The S-statistics.}
    \label{fig:period_xmm}
\end{figure}

\subsection{Spectral Analysis}

We performed spectral analysis of  the  Nustar observation of 4U 0114+65 as a whole, as well as individual sections of the data (Flare A, Quiescent, and Flare B). The extracted spectra were grouped into bins, each with minimum of 20 count using $grppha$ task (exception was the  spectrum  of the quiescent state which  was  grouped into 25 counts per bin to constrain spectral parameters).  We used $XSPEC$ version v12.11.1, simultanously fitting spectra obtained from both focal plane modules (FPMA \& FPMB) in the entire energy band covered by the detectors (i.e. 3-78 keV). We fitted the continuum spectra with the commonly used model for HMXBs, powerlaw modified at high energy by a cutoff ($highecut$) model (White et al. 1983). $Phabs$ model was used to account for absorption below 10 keV. Results of spectral fits are given in table (3) and shown in figure (7). 

Spectral parameters  match  previously obtained values for the system (Hall et al. 2000; Masetti et al. 2006; Farrell et al. 2008). Generally most spectral parameters don't show much change with the change of state. Only photon index $\Gamma$ shows higher value during the quiescent state, and this means that during the quiescent state the spectrum is much softer, however the low statistics in counts above $\sim 20$ keV could have some impact on spectral parameters in such case. The spectra of 4U 0114+65 during the Nustar observation didn't show any significant presence of iron emission line at any time. On the other hand, very tentitive evidence of CRSF absorption presence around $\sim$ 17 keV occasionally was obtained. However  all trial fits adding a $cyclabs$ model didn't improve any of the fits significantly plus giving unreasonally very small width (< 1 keV) of the absorption lines of the fundamental and the first harmonic around $\sim$ 35 keV. The $\chi^{2}$ values improved by 6, 8, 16, and 19 for Quiescent, Flare-A, Flare-B, and total spectrum respectively (all for extra 5 spectral parameters). Although visual inspection show that the fundamental is not even apparent in any of the spectra, deficiency of emission around the expected first and second harmonics at $\sim$ 35 \& 50 keV respectively point to the possibility  for a true  CRSF around $\sim$ 17 keV. Another feature visually apparent in some of the spectra is the presence of excess emission above the continuum in the high energy above 50 keV (i.e. hard-tail). This hard tail is apparent in the overall spectrum and that of flare-A. During the quiescent state only one data point is present above 50 keV, which makes it difficult to firmly associate presence of hard tail in that state. In flare-B there is a complete absence of any emission above 50 keV.

\begin{figure}
	\includegraphics[height=8cm, width=10cm]{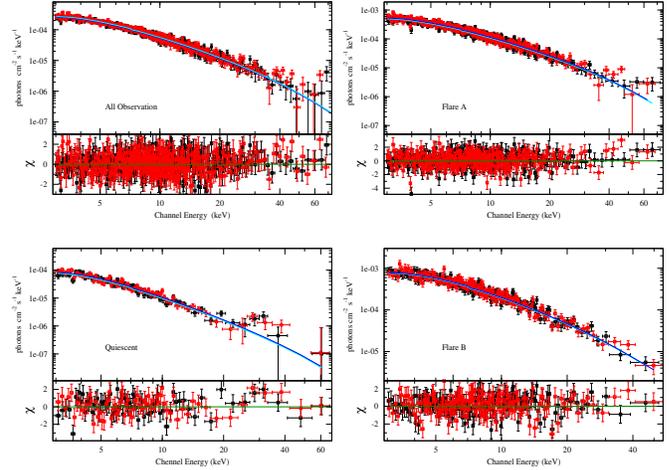}
    \caption{3-78 keV X-ray spectra at different emission states labelled (panels 2, 3, and 4) and the total spectrum (first panel) of Nustar observation FPMA (black points) and FPMB (red point). The delta-chi is shown in the bottom panel of each.}
    \label{fig:period_xmm}
\end{figure}

\section{DISCUSSION AND CONCLUSION}

Stellar winds are not spherically symmetric. In some regions wind is emitted at high speed "high speed stream", while other regions eject material at lower speed "low speed stream". Such variations in wind stream speed along with the rotation of the star sets the conditions needed for the interaction between high/low-speed streams. In the interaction region the material density attains its maximum with speed midway between the speeds of the shocked material bounding the interaction region. The interaction region corotates with the star forming a spiral  pattern  and thus refered to as  Corotating  Interaction Region (CIR). The origin of fast stream and thus CIR formation in low mass stars are regions of open magnetic field "coronal holes". On the other hand in high mass stars dark spots  produce  low density/high speed stream relative to bright spots which produce high density/low speed stream. For a comprehensive discussion of CIR see (Mullan 1984; and Cranmer \& Owocki 1996). 

\vspace{0.3 cm}

During the Nustar observation of 4U 0114+65 two flares occured, one at the begining and the other at the end of the observation. This kind of temporary and sudden increase in X-ray emission that lasts for few ksec has been observed before (Masetti et al. 2006; and Mukherjee \& Paul 2006). These flares should not be confused with those mentioned in earlier work (e.g. Yamauchi et al. 1990; Apparao et al. 1991) which in fact refer to the high state in contrast to the low state of the pulse phase (see figure 3 of Hall 2000 for definition of these states). The flares observed here lasted about 4-6 ksec. Given the system parameters $R_{*} = 37 R_{\odot}$, $M_{*} = 16 M_{\odot}$ (Reig et al. 1996), $P_{orb} = 11.6$ day, and using $1.4 M_{\odot}$ for the neutron star mass, we obtain a neutron star velocity of $v_{ns} = 224 km/s$ and a binary separation of $a = 1.51 R_{*}$. This gives a size of the region during enhanced accretion of about $0.9-1.34 \times 10^{11} cm$. This is comparable to the expected size of CIR $\sim$ $0.1 r$ (Mullan 1984), where $r$ is the distance of CIR from the surface of the normal star.
\vspace{10 mm}

The neutron star in 4U 0114+65 is known to show a general spin up trend which is occasionally interrupted by going through episodes of torque reversal lasting few  hundreds  of days (Sood et al. 2006). The spin up rate seems to be increasing over time: $\dot p$ $\sim$ $-6.2$ $\times$ $10^{-7}$ $s s^{-1}$ from 1986 to 1996; $\sim$ $-8.9$ $\times$ $10^{-7}$ $s s^{-1}$ from 1996 to 2004; $\sim$ $-1.6$ $\times$ $10^{-6}$ $s s^{-1}$ from 2004 to 2006 (Wang 2011 and references therein). Using INTEGRAL/IBIS observations from 2003 to 2008, Wang (2011) found a spin-up rate $\dot p$ $\sim$ $-1.09$ $\times$ $10^{-6}$ $s s^{-1}$.  The spin period evolution of the neutron star in 4U0114+65 system is shown in figure (8).  From the spin periods measurements for the neutron star from XMM-Newton and Nustar observations of the system 4U0114+65, we obtained an average spin-up rate between the observations  $\dot p$ = $1.54$ $\pm$ $0.38$ $\times$ $10^{-6}$ $s$ $s^{-1}$.  The predicted equilibrium period of the neutron star in 4U 0114+65 is $\lesssim$ 26 min (Ikhsanov 2007). With the current value obtained for the spin up rate the neutron star will reach its equilibrium period in   $<$ 200 yr (upper-limit),  or even less if the spin up rate is really  accelerating . It is worth noting that the equilibrium period will be reached in $<$ 100 yr and $<$ 1000 yr for disc-fed and wind-fed accretion, respectively (Ikhsanov 2007). This fact indicates that CIR dominate the accretion torque in the binary system 4U 0114+65. In high mass stars CIR can form very close to the stellar surface and form  a  spiral pattern around the normal star, thus accretion of CIR material could in some times contribute to spin-up torque while other times spin-down torque and what we observe is the mean of these contributions. In high mass X-ray binaries (e.g. GX301-2 Leahy \& Kostka 2008), a spiral density wave pattern can be formed by the excess density in the wind emitted from the stellar surface nearest the L1 Lagrange point. Over the last 40 years these torque contributions favor a mean spin-up trend with some episodes of spin-down/random-walk (Sood et al. 2006; Hu et al. 2017). The main problem here is the negligible probability of observing  the  accreting pulsar during this short lived episode of approching its equilibrium spin period through accretion. Future theoretical work treating X-ray  pulsars  in CIR dominated environment may resolve this problem. 

\vspace{0.3 cm}

In the current work an excess emission above 50 keV is visually noticed in some of the spectra. This hard tail emission is expected from X-ray binaries accreting material through Roche-lobe overflow, where accretion disc corona responsible for comptonization of seed  photons emanating  from the neutron star. This hard X-ray tail from 4U 0114+65 has been occasionally detected before (den Hartog et al. 2006; and Wang 2011). Furthermore, Wang (2011) found anti-correlation between the detection of hard tail and the absorption column density. The higher column density supresses the hard tail and vice versa. The shocks bounding the CIR are an efficient particle accelerator (Mullan 1984), thus provide the necessary heating of material bounding CIR similar to hot corona. This hot material is most probably responsible for the comptonization of X-ray emission from the neutron star. In our spectra, we found excess of emission above 50 keV during flare-A as well as the overall spectrum. During quiescent state only one data point is present above 50 keV and this is not enough to confirm presence of hard tail during quiescent state. In flare-B no signature of hard tails is present at all. In contrary to what was found by Wang (2011), our spectra are all showing high absorption column density. In CIR model association of hard tail with high absorption column density could be realized if the neutron star in immersed in the CIR region and bounded by hot electrons that upscatter the soft photon produced by the neutron star while accreting the bulk material of CIR, this is similar to the case of flare-A here and also  apparent  in the fluctuated activities of X-ray emission pre/post main flare due to accretion of the much diluted material bounding the CIR. Another possibility of such association is the passage of neutron star behind CIR region blocking our line of sight to the X-ray source while allowing scattered emission by the bounding material. During flare-B there  was  no hard X-ray (> 50 keV) emission. This could be understood by the absence of bounding hot material and further confirmed by the absence of pre-flare acctivities. Unfortunately the observation did not extend beyond the end of the second flare. The hot electrons in this case might also had enough time to be cooled by the emission from the first flare (flare-A). CIR model does not require specific association of hard tail with the hydrogen column density. As clear from our results, the anti-correlation between hard tail and column density found by Wang (2011) is most probably accidental anti-correlation.  

\vspace{0.3 cm}

Further spectral evidence for the dominant role of CIR in the system is the previous detection of soft excess despite the high column density (Masetti et al. 2006). Masetti et al. (2006) modeled the soft excess with a blackbody of temperature $\sim$ 0.3 keV with a size $\sim$ 20 times that of the neutron star radius. This soft excess could be in fact self emission from CIR.
\vspace{0.3 cm}

We end our discussion with the most debatable spectral feature from the system 4U 0114+65, namely Cyclotron Resonant Scattering Feature (CRSF). In the current work adding a cyclabs model to the continuum spectra did point to a possible CRSF around $\sim$ 17 keV. The most  probable reason for such detection is the presence of a clear absorption feature around 35 keV which is visually apparent in some of the spectra in figure (7). However, improvement of the fits were not significant to confirm this as a true CRSF signature. Visual inspection of previous spectra from other works show occasionally the presence of such feature downward of 20 keV (the most prominant example is that in figure (11) especially the spectrum of rev. 675 in Wang 2011). Using INTEGRAL/IBIS data of 4U 0114+65 during observational revolutions 262 \& 263, Bonning \& Falanga (2005) detected a possible absorption feature around 22 \& 44 keV. Reanalysis of the data along with other IBIS observations Wang (2011) couldn't confirm the presence of CRSF around 22 keV. Wang (2011) found only a possible hint to features around 40 , 44 ,and 54 keV in some IBIS observations. No other detection of CRSF feature been reported elsewhere. It might be the complex physical environment due to the presence of CIR  hinders  the detection of such feature, similar to that is the case of iron emission line at 6.4 keV which sometimes is detected in the spectra while other times, as the case of the present Nustar observation, no presence of iron line  is found   in any of the emission states (i.e. flares or quiescent).  The main reason for pointing out the possibility of CRSF feature downward of 20 keV is to attract attention for future observations that might confirm such feature as a true signature of CRSF line.

\vspace{0.3 cm}

In summary, we have analysed Nustar and XMM-Newton data to obtain light-curves. From these we find the pulse period and spin-up of the neutron star and observe flares from accretion of CIR material. The Nustar spectra were analyzed,  showing powerlaw with high energy cutoff. No 6.4 keV iron line feature was found, but tentative  hint  for a 17 keV cyclotron feature is seen.

\begin{figure}
	\includegraphics[height=7cm, width=9cm]{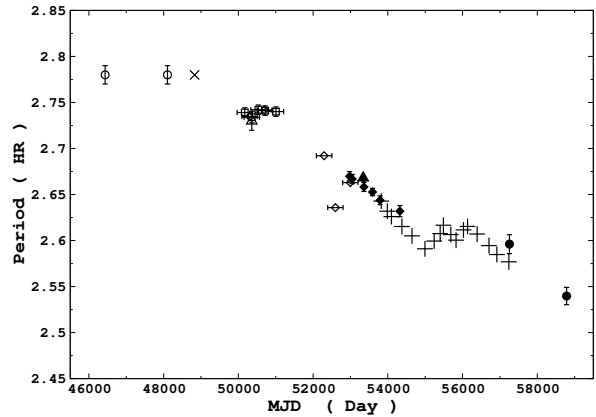}
    \caption{ The spin period evolution of the neutron star in 4U0114+65 system: open circles (Finley et al. 1992), cross (Finley et al. 1994), open squares (Corbet et al. 1999), open triangle (Hall et al. 2000), filled trangle (Bonning \& Falanga 2005), open diamonds (Farrell et al. 2006), filled diamonds (Wang 2011), pluses (few selected points showing the general trend taken from figure "7" of Hu et al. 2017), filled circles (current work).}
    \label{fig:period_xmm}
\end{figure}



\section*{Data Availability}
The data underlying this article will be shared on reasonable request to the corresponding author.
 



\bibliographystyle{mnras}
\bibliography{example} 

\begin{thebibliography}{99}
\bibitem[\protect\citeauthoryear{An et al.}{2014}]{2014SPIE.9144E..1QA} An H., Madsen K.~K., Westergaard N.~J., Boggs S.~E., Christensen F.~E., Craig W.~W., Hailey C.~J., et al., 2014, SPIE, 9144, 91441Q
\bibitem[\protect\citeauthoryear{Apparao, Bisht, \& Singh}{1991}]{1991ApJ...371..772A} Apparao K.~M.~V., Bisht P., Singh K.~P., 1991, ApJ, 371, 772
\bibitem[\protect\citeauthoryear{Aschenbach et al.}{2000}]{2000SPIE.4012..731A} Aschenbach B., Briel U.~G., Haberl F., Braeuninger H.~W., Burkert W., Oppitz A., Gondoin P., et al., 2000, SPIE, 4012, 731
\bibitem[\protect\citeauthoryear{Ashok et al.}{2006}]{2006AdSpR..38.2777A} Ashok N.~M., Manchanda R.~K., Banerjee D.~P.~K., Farrell S., Sood R.~K., 2006, AdSpR, 38, 2777
\bibitem[\protect\citeauthoryear{Bonning \& Falanga}{2005}]{2005A&A...436L..31B} Bonning E.~W., Falanga M., 2005, A\&A, 436, L31
\bibitem[\protect\citeauthoryear{Corbet, Finley, \& Peele}{1999}]{1999ApJ...511..876C} Corbet R.~H.~D., Finley J.~P., Peele A.~G., 1999, ApJ, 511, 876
\bibitem[\protect\citeauthoryear{Crampton, Hutchings, \& Cowley}{1985}]{1985ApJ...299..839C} Crampton D., Hutchings J.~B., Cowley A.~P., 1985, ApJ, 299, 839
\bibitem[\protect\citeauthoryear{Cranmer \& Owocki}{1996}]{1996ApJ...462..469C} Cranmer S.~R., Owocki S.~P., 1996, ApJ, 462, 469
\bibitem[\protect\citeauthoryear{den Hartog et al.}{2006}]{2006A&A...451..587D} den Hartog P.~R., Hermsen W., Kuiper L., Vink J., in't Zand J.~J.~M., Collmar W., 2006, A\&A, 451, 587
\bibitem[\protect\citeauthoryear{Dower \& Kelley}{1977}]{1977IAUC.3144....2D} Dower R., Kelley R., 1977, IAUC, 3144, 2
\bibitem[\protect\citeauthoryear{Farrell, Sood, \& O'Neill}{2006}]{2006MNRAS.367.1457F} Farrell S.~A., Sood R.~K., O'Neill P.~M., 2006, MNRAS, 367, 1457
\bibitem[\protect\citeauthoryear{Farrell et al.}{2008}]{2008MNRAS.389..608F} Farrell S.~A., Sood R.~K., O'Neill P.~M., Dieters S., 2008, MNRAS, 389, 608
\bibitem[\protect\citeauthoryear{Finley, Belloni, \& Cassinelli}{1992}]{1992A&A...262L..25F} Finley J.~P., Belloni T., Cassinelli J.~P., 1992, A\&A, 262, L25
\bibitem[\protect\citeauthoryear{Finley, Taylor, \& Belloni}{1994}]{1994ApJ...429..356F} Finley J.~P., Taylor M., Belloni T., 1994, ApJ, 429, 356
\bibitem[\protect\citeauthoryear{Grundstrom et al.}{2007}]{2007ApJ...656..431G} Grundstrom E.~D., Blair J.~L., Gies D.~R., Huang W., McSwain M.~V., Raghavan D., Riddle R.~L., et al., 2007, ApJ, 656, 431
\bibitem[\protect\citeauthoryear{Hall et al.}{2000}]{2000ApJ...536..450H} Hall T.~A., Finley J.~P., Corbet R.~H.~D., Thomas R.~C., 2000, ApJ, 536, 450
\bibitem[\protect\citeauthoryear{Harrison et al.}{2013}]{2013ApJ...770..103H} Harrison F.~A., Craig W.~W., Christensen F.~E., Hailey C.~J., Zhang W.~W., Boggs S.~E., Stern D., et al., 2013, ApJ, 770, 103
\bibitem[\protect\citeauthoryear{Hu et al.}{2017}]{2017ApJ...844...16H} Hu C.-P., Chou Y., Ng C.-Y., Lin L.~C.-C., Yen D.~C.-C., 2017, ApJ, 844, 16
\bibitem[\protect\citeauthoryear{Ikhsanov}{2007}]{2007MNRAS.375..698I} Ikhsanov N.~R., 2007, MNRAS, 375, 698
\bibitem[\protect\citeauthoryear{Jansen et al.}{2001}]{2001A&A...365L...1J} Jansen F., Lumb D., Altieri B., Clavel J., Ehle M., Erd C., Gabriel C., et al., 2001, A\&A, 365, L1
\bibitem[\protect\citeauthoryear{Koenigsberger et al.}{1983}]{1983ApJ...268..782K} Koenigsberger G., Swank J.~H., Szymkowiak A.~E., White N.~E., 1983, ApJ, 268, 782
\bibitem[\protect\citeauthoryear{Kotze \& Charles}{2012}]{2012MNRAS.420.1575K} Kotze M.~M., Charles P.~A., 2012, MNRAS, 420, 1575 
\bibitem[\protect\citeauthoryear{Leahy et al.}{1983}]{1983ApJ...266..160L} Leahy D.~A., Darbro W., Elsner R.~F., Weisskopf M.~C., Sutherland P.~G., Kahn S., Grindlay J.~E., 1983, ApJ, 266, 160
\bibitem[\protect\citeauthoryear{Leahy}{1987}]{1987A&A...180..275L} Leahy D.~A., 1987, A\&A, 180, 275
\bibitem[\protect\citeauthoryear{Leahy \& Kostka}{2008}]{2008MNRAS.384..747L} Leahy D.~A., Kostka M., 2008, MNRAS, 384, 747
\bibitem[\protect\citeauthoryear{Li \& van den Heuvel}{1999}]{1999ApJ...513L..45L} Li X.-D., van den Heuvel E.~P.~J., 1999, ApJL, 513, L45
\bibitem[\protect\citeauthoryear{Lomb}{1976}]{1976Ap&SS..39..447L} Lomb N.~R., 1976, Ap\&SS, 39, 447
\bibitem[\protect\citeauthoryear{Margon \& Bradt}{1977}]{1977IAUC.3144....2D} Margon B., Bradt H., 1977, IAUC, 3144, 2
\bibitem[\protect\citeauthoryear{Masetti et al.}{2006}]{2006A&A...445..653M} Masetti N., Orlandini M., dal Fiume D., del Sordo S., Amati L., Frontera F., Palazzi E., et al., 2006, A\&A, 445, 653
\bibitem[\protect\citeauthoryear{Mukherjee \& Paul}{2006}]{2006JApA...27...37M} Mukherjee U., Paul B., 2006, JApA, 27, 37
\bibitem[\protect\citeauthoryear{Mullan}{1984}]{1984ApJ...283..303M} Mullan D.~J., 1984, ApJ, 283, 303
\bibitem[\protect\citeauthoryear{Reig et al.}{1996}]{1996A&A...311..879R} Reig P., Chakrabarty D., Coe M.~J., Fabregat J., Negueruela I., Prince T.~A., Roche P., et al., 1996, A\&A, 311, 879
\bibitem[\protect\citeauthoryear{Sanjurjo-Ferrr{\'\i}n et al.}{2017}]{2017A&A...606A.145S} Sanjurjo-Ferrr{\'\i}n G., Torrej{\'o}n J.~M., Postnov K., Oskinova L., Rodes-Roca J.~J., Bernabeu G., 2017, A\&A, 606, A145
\bibitem[\protect\citeauthoryear{Scargle}{1982}]{1982ApJ...263..835S} Scargle J.~D., 1982, ApJ, 263, 835
\bibitem[\protect\citeauthoryear{Sood et al.}{2006}]{2006AdSpR..38.2779S} Sood R., Farrell S., O'Neill P., Manchanda R., Ashok N.~M., 2006, AdSpR, 38, 2779
\bibitem[\protect\citeauthoryear{Stellingwerf}{1978}]{1978ApJ...224..953S} Stellingwerf R.~F., 1978, ApJ, 224, 953
\bibitem[\protect\citeauthoryear{Wang}{2011}]{2011MNRAS.413.1083W} Wang W., 2011, MNRAS, 413, 1083
\bibitem[\protect\citeauthoryear{Wen et al.}{2006}]{2006ApJS..163..372W} Wen L., Levine A.~M., Corbet R.~H.~D., Bradt H.~V., 2006, ApJS, 163, 372
\bibitem[\protect\citeauthoryear{White, Swank, \& Holt}{1983}]{1983ApJ...270..711W} White N.~E., Swank J.~H., Holt S.~S., 1983, ApJ, 270, 711
\bibitem[\protect\citeauthoryear{Yamauchi et al.}{1990}]{1990PASJ...42L..53Y} Yamauchi S., Asaoka I., Kawada M., Koyama K., Tawara Y., 1990, PASJ, 42, L53

\end{thebibliography}








\bsp	
\label{lastpage}
\end{document}